\documentclass[aps,prl,twocolumn,superscriptaddress,longbibliography]{revtex4-2}

\usepackage{textcomp}
\usepackage{amsmath}
\usepackage{cancel}
\usepackage{graphicx}
\usepackage{xspace}
\usepackage{amsfonts}
\usepackage{color}

\makeatletter

\definecolor{red}{rgb}{0.75,0,0}
\definecolor{blue}{rgb}{0,0,0.75}
\definecolor{green}{rgb}{0,0.5,0}
\definecolor{orange-red}{RGB}{255,69,0}

\newcommand{\ave}[1]{\left\langle #1 \right\rangle}

\usepackage{dsfont}

\newcommand{\canetset}[1]{{\mathchoice {\hbox{$\sf\textstyle #1\kern-0.4em #1$}}
{\hbox{$\sf\textstyle #1\kern-0.4em #1$}}
{\hbox{$\sf\scriptstyle #1\kern-0.3em #1$}}
{\hbox{$\sf\scriptscriptstyle #1\kern-0.2em #1$}}}}

\def\nbZ{{\mathchoice {\hbox{$\sf\textstyle Z\kern-0.4em Z$}}
{\hbox{$\sf\textstyle Z\kern-0.4em Z$}}
{\hbox{$\sf\scriptstyle Z\kern-0.3em Z$}}
{\hbox{$\sf\scriptscriptstyle Z\kern-0.2em Z$}}}}
\usepackage{bm}

\newlength \standardfigwidth
\newcounter{exercise}
{\addtocounter{exercise}{1}\begin{center}\begin{minipage}{0.8\linewidth}\textbf{Exercise
\arabic{exercise}:}\begin{itshape}}% begin code
{\end{itshape}\end{minipage}\end{center}}%                    end code

\makeatletter
\newcommand{\creat}[3][]{\@ifempty{#1}{#2^{\dagger}}{\left(#2^{\dagger}\right)^{#1}}\@ifempty{#3}{}{\!(#3)}}

\newcommand{\creatDoi}[3][]{\@ifempty{#1}{\tilde{#2}}{\left(\tilde{#2}\right)^{#1}}\@ifempty{#3}{}{(#3)}}

\newcommand{\annih}[3][]{#2\@ifempty{#1}{}{^{#1}}\@ifempty{#3}{}{(#3)}}

\makeatother

% ...
\newlength{\bibmarkkeyAleft}

\newlength{\bibmarkkeyBleft}

\newlength{\bibmarkkeyCleft}

\newlength{\bibmarkkeyDleft}

\newcommand{\distance}{q}

\begin{document}

\title{An Informational Route to Negative Mobility}

\author{Ziluo Zhang}
%\thanks{Corresponding author: zhyou@xmu.edu.cn}
\affiliation{Fujian Provincial Key Laboratory for Soft Functional Materials Research,
Research Institute for Biomimetics and Soft Matter, Department of
Physics, Xiamen University, Xiamen, Fujian 361005, China}

\author{Shigeyuki Komura}
\thanks{Corresponding author: komura@wiucas.ac.cn}
\affiliation{Zhejiang Key Laboratory of Soft Matter Biomedical Materials, Wenzhou Institute, University of Chinese Academy of Sciences, Wenzhou 325001, China}

\author{Zhihong You}
\thanks{Corresponding author: zhyou@xmu.edu.cn}
\affiliation{Fujian Provincial Key Laboratory for Soft Functional Materials Research,
Research Institute for Biomimetics and Soft Matter, Department of
Physics, Xiamen University, Xiamen, Fujian 361005, China}

\date{\today}
%图1c->图1b, 画f<0
%图c 小f y轴moblity

%图4a->图2

%图3d 散点
%=====================================================
\begin{abstract}
Mobility links an applied force to the resulting motion and is generally positive near equilibrium. Far from equilibrium, however, internal energy input can reverse this response. Here we show that information feedback provides a distinct route to negative mobility. We consider an overdamped dimer consisting of a run-and-tumble particle coupled by a spring to a passive Brownian particle. Information enters through periodic measurements of the relative displacement, which are processed to reset the active polarity. Although the unloaded dimer has no net drift, an applied force biases its internal configuration, and an information–mechanical feedback amplifies and converts this bias into active propulsion against the force, producing negative mobility.
%Although the dimer has no drift without a load, an applied force biases its internal configuration, and feedback converts this bias into active propulsion against the force. The resulting propulsion further strengthens the configurational bias, leading to negative mobility. 
We develop an analytical theory that captures this mechanism and yields a feedback-gain criterion for response reversal. Including the information-processing cost reveals a tradeoff: rapid feedback enhances reverse transport but incurs a growing informational cost, yielding an optimal finite feedback rate for information-inclusive efficiency.
%These results show that information can directly modify the mechanical response of active systems by controlling how active propulsion responds to external forces.
These results show that information can reshape nonequilibrium response by controlling how internally supplied energy is converted into macroscopic transport.
\end{abstract}

\maketitle

% ========================================================
Mobility, defined as the response of a system to an applied force, is a fundamental measure of nonequilibrium transport. Near equilibrium, dissipation generally leads to a positive mobility, so that a force drives motion along its direction \cite{CallenWelton1951,Kubo1957}. Far from equilibrium, however, internal energy consumption can reshape this relation, allowing the response to become history dependent, nonreciprocal, and even reverse its sign \cite{Reimann2002,HanggiMarchesoni2009,DavisProesmansFodor2024}. Such absolute negative mobility, where an arbitrarily weak force induces motion opposite to its direction, has been realized through diverse mechanisms, including spatial or geometric asymmetry \cite{HanggiEtAl2010,DuMei2012}, inertial dynamics \cite{EichhornReimannHanggi2002,MachuraEtAl2007,LuoZengAi2020}, confinement \cite{GoldfarbBurov2026}, active many-body effects \cite{Rizkallah2023}, and interparticle interactions \cite{JanuszewskiLuczka2011,SpeerEtAl2012,LiuWang2025,LiuLuoWang2026}. These studies reveal diverse routes to response reversal, yet how internally supplied nonequilibrium energy is organized into a macroscopic response under external forcing remains unclear.

Information provides a distinct route for controlling nonequilibrium systems by coupling measurements to feedback-driven actions \cite{TouchetteLloyd2000,SagawaUeda2008,SagawaUeda2010}. In stochastic thermodynamics, information acquired through measurement can serve as an operational resource, enabling work extraction and modifying fundamental energetic constraints \cite{Landauer1961,Bennett1982,SagawaUeda2009,ToyabeEtAl2010,BerutEtAl2012,HorowitzEsposito2014,ParrondoHorowitzSagawa2015,PaneruEtAl2018}. Active matter offers a natural extension of this framework because active particles continuously consume energy, while feedback can determine how this energy is deployed rather than merely modify fluctuations \cite{BechingerEtAl2016,MalgarettiStark2022,CocconiKnightRoberts2023,DavisProesmansFodor2024,SchuettlerEtAl2025,GarciaMillanEtAl2025}. Recent studies have used information to steer active particles, regulate their dynamics, and construct information-powered engines \cite{MijalkovEtAl2016,ColabreseEtAl2017,SchneiderStark2019,FernandezRodriguezEtAl2020,FranzlCichos2020,MuinosLandinEtAl2021,FalkEtAl2021,MalgarettiStark2022,GoerlichEtAl2022,CocconiKnightRoberts2023,BaldovinGueryOdelinTrizac2023,CocconiChen2024,HouZhangKomura2025,SchuettlerEtAl2025,GarciaMillanEtAl2025}. Measurements of particle configurations have also enabled information-mediated interactions and effective active couplings \cite{KhadkaEtAl2018,BauerleEtAl2018}. These studies show that information can control active motion and modify nonequilibrium dynamics. However, it remains largely unexplored how information can be used to direct internally supplied active energy to reshape the response to an external force.

Here we address this question by introducing an information-mediated active dimer in which the internal mechanical coordinate itself serves as the information channel. The system consists of a one-dimensional run-and-tumble particle harmonically coupled to a Brownian particle, with informational feedback that measures their relative displacement and adjusts the active polarity accordingly. Because the feedback preserves inversion symmetry, the unloaded system remains unbiased with no net drift.
%Under an applied load, the internal separation becomes biased, and feedback converts this bias into an active response opposing the load, resulting in negative mobility. 
Under an applied load, the internal separation becomes biased, and informational feedback converts this bias into an active polarity opposing the load while further amplifying the configurational bias, forming a self-reinforcing information-mechanical coupling that produces negative mobility. The rectifying element is therefore not encoded in a pre-existing mechanical landscape, but emerges dynamically as an informational asymmetry. We establish an exact relation linking biased internal configuration, feedback-induced active polarity, and transport, providing a generic picture of how information reshapes the mechanical response. Finally, accounting for information-processing cost reveals a tradeoff between mechanical output and information-inclusive efficiency, yielding an optimal finite feedback rate that balances stronger reverse transport against the growing cost of information processing. These results show that information can direct active energy by coupling internal state to active dynamics, thereby reshaping the response to external forces.

\begin{figure}
    \centering
    \includegraphics[width=0.9\linewidth]{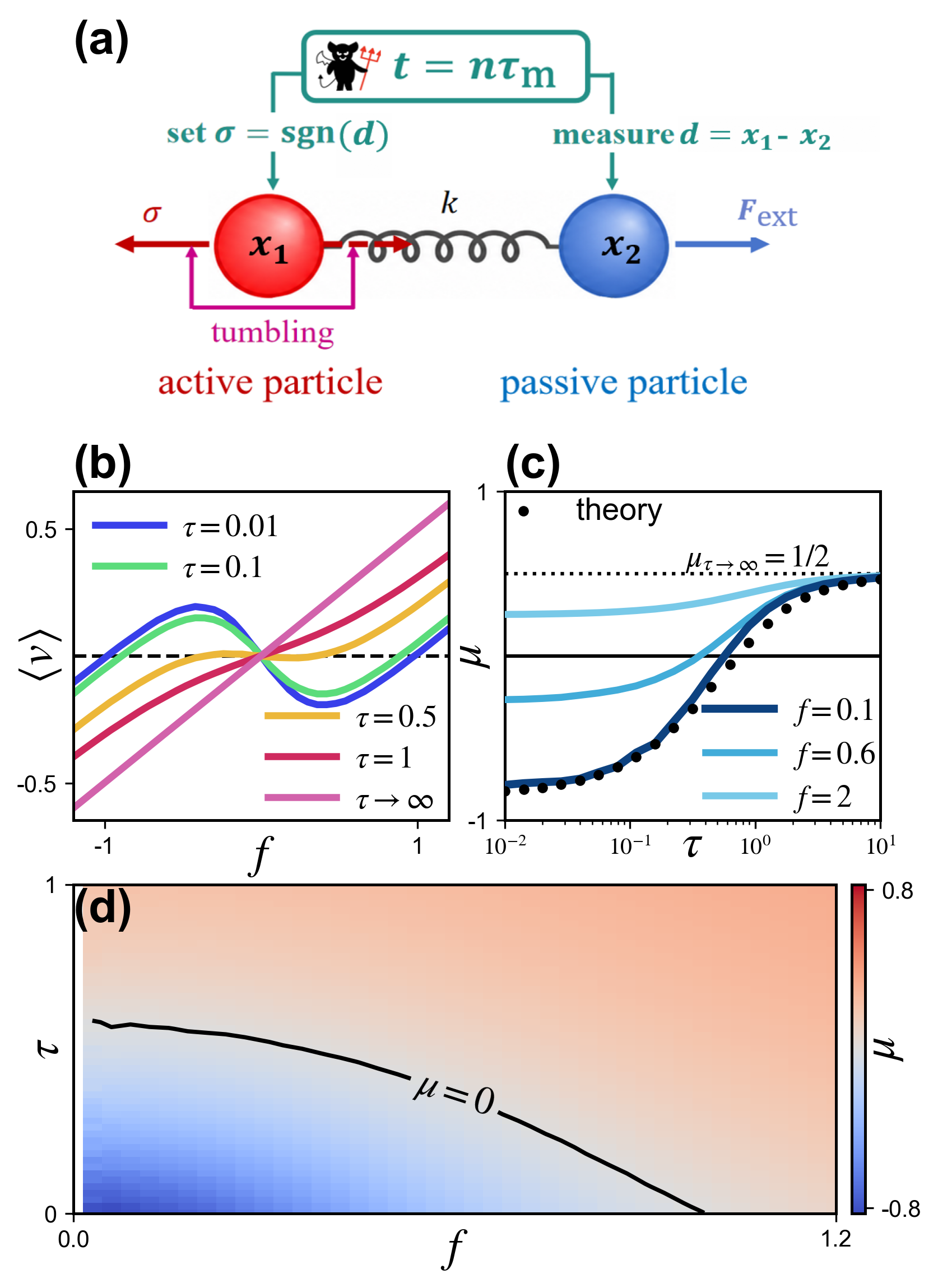}
   \caption{
Model and transport response of the feedback-controlled active--passive dimer.
(a) Schematic of the measurement-and-feedback protocol. 
(b) Steady-state mean velocity $\langle v\rangle$ as a function of the dimensionless force $f$ for different measurement intervals $\tau$. (c) Effective mobility $\mu$ versus $\tau$ for various dimensionless force $f$. Symbols indicate the weak-load analytical prediction from Eq.~\eqref{eq:mu0}.
(d) Mobility diagram in the $(f,\tau)$ plane.
}
    \label{fig:fig1}
\end{figure}

% =====================================================
\textit{Model.}
We consider the minimal one-dimensional active-passive system illustrated in Fig.~\ref{fig:fig1}(a). Active-passive dimers, consisting of a self-propelled particle coupled to a passive cargo or load, have been studied in the context of active transport and taxis \cite{VuijkEtAl2021,MuzzedduEtAl2023,ValechaEtAl2025}. Here, the active component is a run-and-tumble particle with polarity-switching rate $\alpha$ at position $x_1$, harmonically coupled to a passive Brownian particle at position $x_2$.
%We consider the minimal one-dimensional active--passive dimer illustrated in Fig.~\ref{fig:fig1}(a). It consists of a run-and-tumble particle at position $x_1(t)$, harmonically coupled to a Brownian particle at position $x_2(t)$. 
In the overdamped limit, their dynamics are governed by
\begin{subequations}
\label{eq:langevin}
\begin{align}
\gamma \dot{x}_1 &= \gamma u_0 \sigma -k(x_1-x_2)
+\xi_1(t),
\label{eq:langevin_active}
\\
\gamma \dot{x}_2 &= F_{\mathrm{ext}} -k(x_2-x_1) +\xi_2(t),
\label{eq:langevin_passive}
\end{align}
\end{subequations}
where $\dot{x}=dx/dt$, $\gamma$ is the  drag coefficient for both particles, $u_0$ is the self-propulsion speed, $\sigma(t)=\pm1$ is the instantaneous active polarity, and $k$ is the spring stiffness. The active polarity $\sigma$ reverses at Poisson rate $\alpha$. A constant external force $F_{\mathrm{ext}}$ acts only on the passive particle. Both particles are coupled to thermal reservoirs at temperature $T$. The corresponding thermal forces are independent Gaussian white noises with zero mean and correlations $\langle \xi_i(t)\xi_j(t')\rangle = 2\gamma k_{\mathrm{B}} T\,\delta_{ij}\delta(t-t')$.
Information processing is implemented through periodic measurement and feedback at times $t_n=n\tau_{\mathrm{m}}$, where $n$ is the number of the measurement event, $\tau_{\mathrm{m}}$ is the measurement interval. At each update, the controller measures the signed separation $d=x_1-x_2$, which specifies the internal polarization of the dimer, and resets the active polarity according to $\sigma(t_n)=\operatorname{sgn}\left[d(t_n)\right]$. The active particle is thus reoriented away from its passive partner: $\sigma=+1$ when $d>0$ and $\sigma=-1$ when $d<0$. The rule does not favor either direction: reversing both $d$ and $\sigma$ simply maps the dynamics onto its mirror image. Between successive measurements, the particle resumes its intrinsic run-and-tumble dynamics, reversing its polarization as $\sigma\rightarrow-\sigma$ according to rate $\alpha$.

We render the model dimensionless by scaling time with the tumbling time $\alpha^{-1}$, length with the mean ru

.n length $\ell_0=u_0/\alpha$, and force with the self-propulsion force $\gamma u_0$. For convenience, we introduce the dimensionless variables: particle positions $\tilde{x}_i=x_i/\ell_0$, relative separation $q=d/\ell_0$, center-of-mass velocity $v=(\dot{x}_1+\dot{x}_2)/(2u_0)$, measurement interval $\tau=\alpha\tau_{\mathrm m}$, applied force $f=F_{\mathrm{ext}}/(\gamma u_0)$, temperature $\mathcal{T}=\alpha k_{\mathrm B}T/(\gamma u_0^2)$, and spring stiffness $\kappa=k/(\gamma\alpha)$.
Details of numerical simulation are presented in the Supplemental Material Sec.~S1. Unless otherwise stated, the simulations use $\kappa=0.5$ and $\mathcal{T}=0.2$, with $f$ and $\tau$ varied.

% =====================================================
\textit{Negative mobility.}
In the absence of an external force, the composite exhibits no net transport: inversion symmetry renders the two propagation directions statistically equivalent, regardless of whether information feedback is present. Accordingly, the mean velocity vanishes at zero load, $\langle v\rangle=0$, as shown in Fig.~\ref{fig:fig1}(b). Here, $\langle\cdot\rangle$ indicates the ensemble average of the observable. When a force is applied, the information-free dimer displays the conventional mechanical response: it drifts along the force, with a mean velocity that increases linearly with the applied load and a positive effective mobility $\mu=\langle v\rangle/f$ [magenta curve in Fig.~\ref{fig:fig1}(b)].

Information feedback qualitatively reshapes the force response. When the measurement interval greatly exceeds the intrinsic run time, $\tau\gg 1$, the influence of each update decays well before the next measurement, and the dimer behaves similarly to its information-free counterpart. As the measurements become more frequent, the mobility decreases progressively [red and orange curves in Fig.~\ref{fig:fig1}(b)]. For $\tau$ sufficiently below unity, the mobility at weak loads becomes negative, whereas at larger loads the mobility is reduced but remains positive [Fig.~\ref{fig:fig1}(b) and (c)]. In the rapid-measurement limit, $\tau\ll 1$, the force–velocity relation converges to the limiting blue curve in Fig.~\ref{fig:fig1}(b), indicating saturation of the mobility. Figure~\ref{fig:fig1}(c) summarizes this crossover by showing the mobility as a function of the measurement interval: a clear transition from positive to negative mobility emerges at small loads, while at larger loads the feedback weakens the response without reversing its sign. Consistently, the mobility map in Fig.~\ref{fig:fig1}(d) reveals a broad negative-mobility regime, which progressively narrows and ultimately vanishes as the measurement interval increases. These results reveal a crossover between a feedback-dominated regime, in which information reverses the weak-load response, and a force-dominated regime, in which the ordinary positive response is restored. The saturation at short measurement intervals further indicates that the response is limited not by the measurement rate itself, but by the bounded active polarity $-1\leq\ave{\sigma}\leq1$.

% =====================================================
\textit{Information-mediated reversal mechanism.}
The microscopic origin of the negative response lies in an self-reinforcing information-mechanics coupling between the dimer's internal state and active polarity, mediated by the external force. This mechanism is most transparently revealed by the relation between the transport response and the active polarity. Since the spring force is internal, it cancels from the center-of-mass dynamics, and the two bound particles share the same long-time average velocity. The force balance of the composite therefore gives the exact relation 
\begin{equation}
\label{eq:v}
\langle v\rangle=\frac{1}{2}(f+\langle\sigma\rangle).
\end{equation}
The velocity thus contains two distinct contributions: the bare mechanical drift $f/2$ induced by the external force and the active contribution $\langle\sigma\rangle/2$ due to self-propulsion. For a positive load, response reversal occurs when $\langle\sigma\rangle<-f$, where the information-mediated active response overwhelms the direct mechanical response. At larger forces, however, the direct mechanical contribution eventually dominates because the active polarity is bounded. The polarity curve therefore intersects the zero-velocity boundary at a finite stall force.

\begin{figure}
    \centering
    \includegraphics[width=0.9\linewidth]{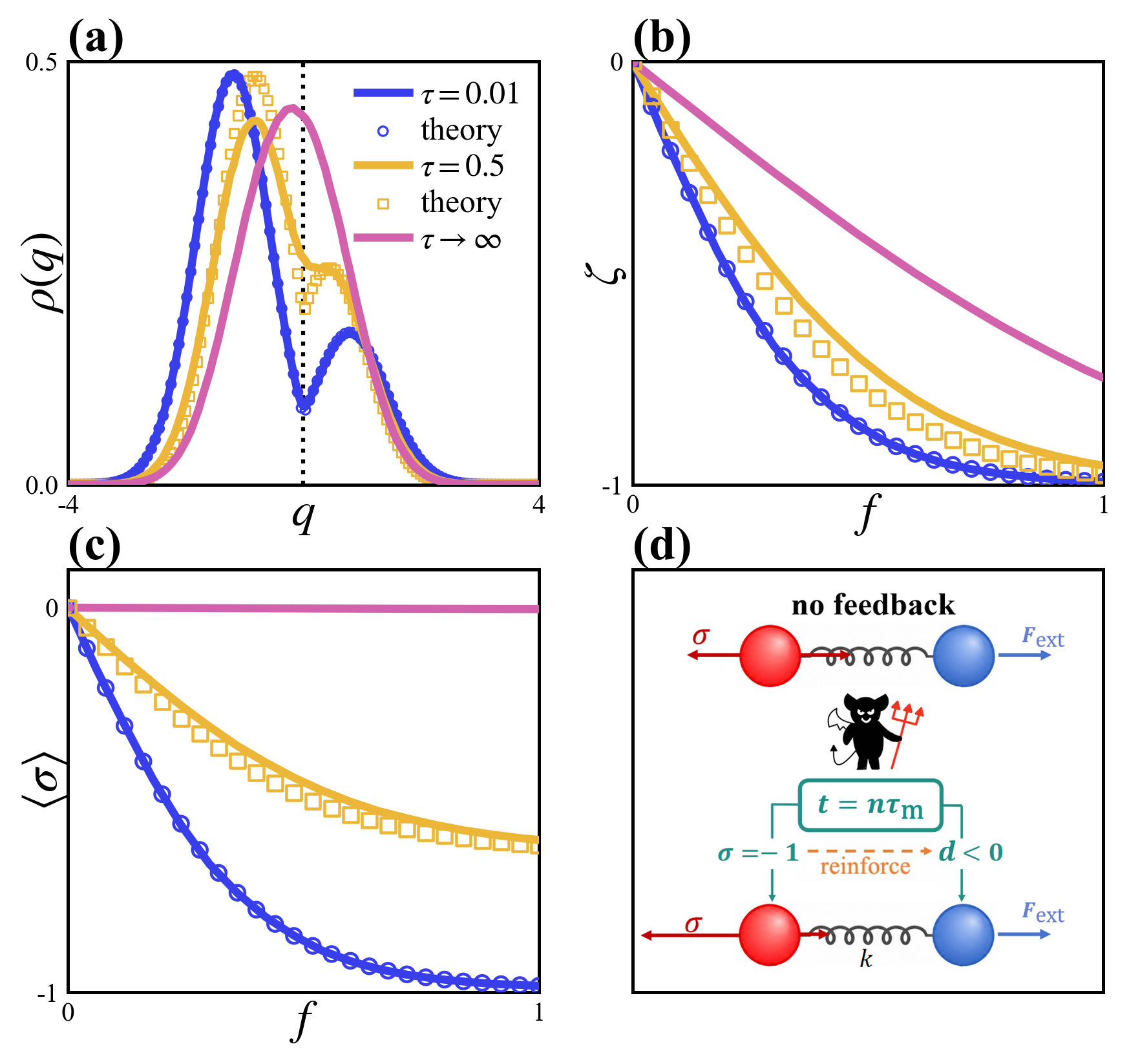}
   \caption{Self-reinforcing information–mechanical feedback from configurational bias to active response.
   (a) Steady-state distribution $\rho(\distance)$ of the relative coordinate at $f=0.2$ for representative measurement intervals. 
   (b) Internal polarity $\zeta$ as a function of $f$.
   (c) Mean active polarity $\ave{\sigma}$ as a function of the external force $f$ for different $\tau$. (d) Schematic comparison between information-free dynamics and the measurement-and-feedback protocol. Solid lines show simulations, while symbols denote the analytical prediction.
}
    \label{fig:fig2}
\end{figure}

To identify the origin of the nonzero active polarity $\langle\sigma\rangle$, we characterize the internal state by the distribution $\rho(q)$ of the relative separation and an internal polarity $\zeta=\langle\operatorname{sgn}(q)\rangle$. At zero load, inversion symmetry requires $\zeta=0$, so both internal and active polarities vanish. A positive force on the passive particle shifts it to the right relative to the active particle, favoring $\distance<0$ and thereby producing a negative internal polarity [Figs.~\ref{fig:fig2}(a) and \ref{fig:fig2}(b)]. Without feedback, this record remains dynamically inert because the run-and-tumble orientation is independent of the internal state, leaving $\langle\sigma\rangle=0$ and preserving the ordinary positive response [purple curve in Fig.~\ref{fig:fig2}(c) and upper panel of Fig.~\ref{fig:fig2}(d)]. 
Feedback processes this internal record by resetting the active polarization according to $\sigma(t_n)=\operatorname{sgn}[\distance(t_n)]$. Importantly, the resulting propulsion not only acts on the measured internal state but also reinforces it: by driving the active particle away from its passive partner, it increases the magnitude of the existing separation and strengthens the corresponding configurational bias [lower panel of Fig.~\ref{fig:fig2}(d)]. The feedback therefore establishes a self-reinforcing information-mechanics coupling: the measured configuration sets the active polarity, and the resulting propulsion in turn strengthens the same configurational bias. Consequently, both $\zeta$ and $\langle\sigma\rangle$ become more negative with increasing load or measurement frequency [Figs.~\ref{fig:fig2}(a)--\ref{fig:fig2}(c)]. Intrinsic tumbling opposes this amplification by progressively erasing the measurement-induced active polarity, yielding $\langle\sigma\rangle=\chi\zeta$, where $\chi=[1-e^{-2\tau}]/(2\tau)$ quantifies the polarity retained between successive measurements (see Sec.~S2A).

The information-mediated response can be quantified analytically in the rapid-measurement regime $\tau\ll1$ (see Sec.~S2). In this limit, frequent resetting slaves the active polarity to the instantaneous internal state, allowing the relative separation to be described by the Boltzmann-like distribution $\rho(\distance)\propto\text{exp}[(-\kappa\distance^2-f\distance+\chi|\distance|)/(2\mathcal T)]$ [blue and yellow dots in Fig.~\ref{fig:fig2}(a)] \cite{Risken:1989}. The external force biases the distribution toward $\distance<0$, while feedback amplifies this bias by separating the two configurational sectors. Integrating over $\distance$ gives the internal polarization $\zeta=[g(\chi-f)-g(\chi+f)]/[g(\chi-f)+g(\chi+f)]$, where $g(x)=\int_0^\infty \text{exp}[(-\kappa\distance^2+x\distance)/(2\mathcal T)]d\distance$, in good agreement with simulations [blue and yellow dots in Fig.~\ref{fig:fig2}(b)]. %For weak loads, $\zeta=-\lambda f+O(f^3)$, with $\lambda(\chi)=g'(\chi)/g(\chi)$ quantifying how sensitively the binary internal state responds to a small force. Information feedback therefore amplifies the force sensitivity of the internal state by a factor $\lambda(\chi)/\lambda(0)\simeq\chi\sqrt{\pi/(8\kappa\mathcal T)}$, reaching a maximum of $\sqrt{\pi/(8\kappa\mathcal T)}$ as $\tau\to0$, which can greatly exceed unity at low temperature.
For weak loads, $\zeta=-\lambda f+O(f^3)$, with $\lambda(\chi)=g'(\chi)/g(\chi)$ quantifying how sensitively the internal state responds to a small force. The information-free relative dynamics exactly map onto those of a harmonically confined run-and-tumble particle, giving the reference susceptibility $\lambda_0=\operatorname{erf}[1/\sqrt{4\mathcal T}]$ for the present case $\kappa=1/2$ (Sec.~S2C) \cite{GarciaMillanPruessner2021}. Within the rapid-feedback theory, the configurational response is therefore amplified by the information–mechanical feedback by a factor $\lambda(\chi)/\lambda_0$. The amplification reaches its maximum $\lambda(1)/\lambda_0$ as $\tau\to0$, which can greatly exceed unity at low temperature.

The informational nature of $\lambda$ follows by introducing the entropy deficit $\Delta h=h(1/2)-h(p)$, where $h(p)=-p\ln p-(1-p)\ln(1-p)$ is the  binary Shannon entropy and $p=(1+\zeta)/2$ is the probability of measuring $\distance>0$. The reference value $h(1/2)=\ln2$ corresponds to an unbiased configuration with equally probable signs of $\distance$. Hence, $\Delta h$ measures how much the applied force reduces the uncertainty of the binary configurational state. At weak loads, $\Delta h=\zeta^2/2+O(\zeta^4)$, so that $\lambda\approx\sqrt{2\Delta h}/|f|$. Thus, $\lambda$ acts as an information-encoding susceptibility: it quantifies how efficiently a mechanical perturbation is converted into a statistically distinguishable internal record. Feedback then converts this encoded bias into the active response, $\langle\sigma\rangle=-\chi\lambda f+O(f^3)$, yielding the weak-load mobility (see Secs.~S2D and S2E)
\begin{equation}
\label{eq:mu0}
\mu_0=\frac{1}{2}(1-\chi\lambda),
\end{equation}
in quantitative agreement with simulations [Fig.~\ref{fig:fig1}(c)]. This expression separates the information-mediated response into two ingredients: how strongly perturbations are encoded into internal states and how faithfully these records are converted into action. Their product $\chi\lambda$ is the feedback gain, which suppresses the bare response for $\chi\lambda<1$ and reverses it through active overscreening once $\chi\lambda>1$.

\begin{figure}
    \centering
    \includegraphics[width=1\linewidth]{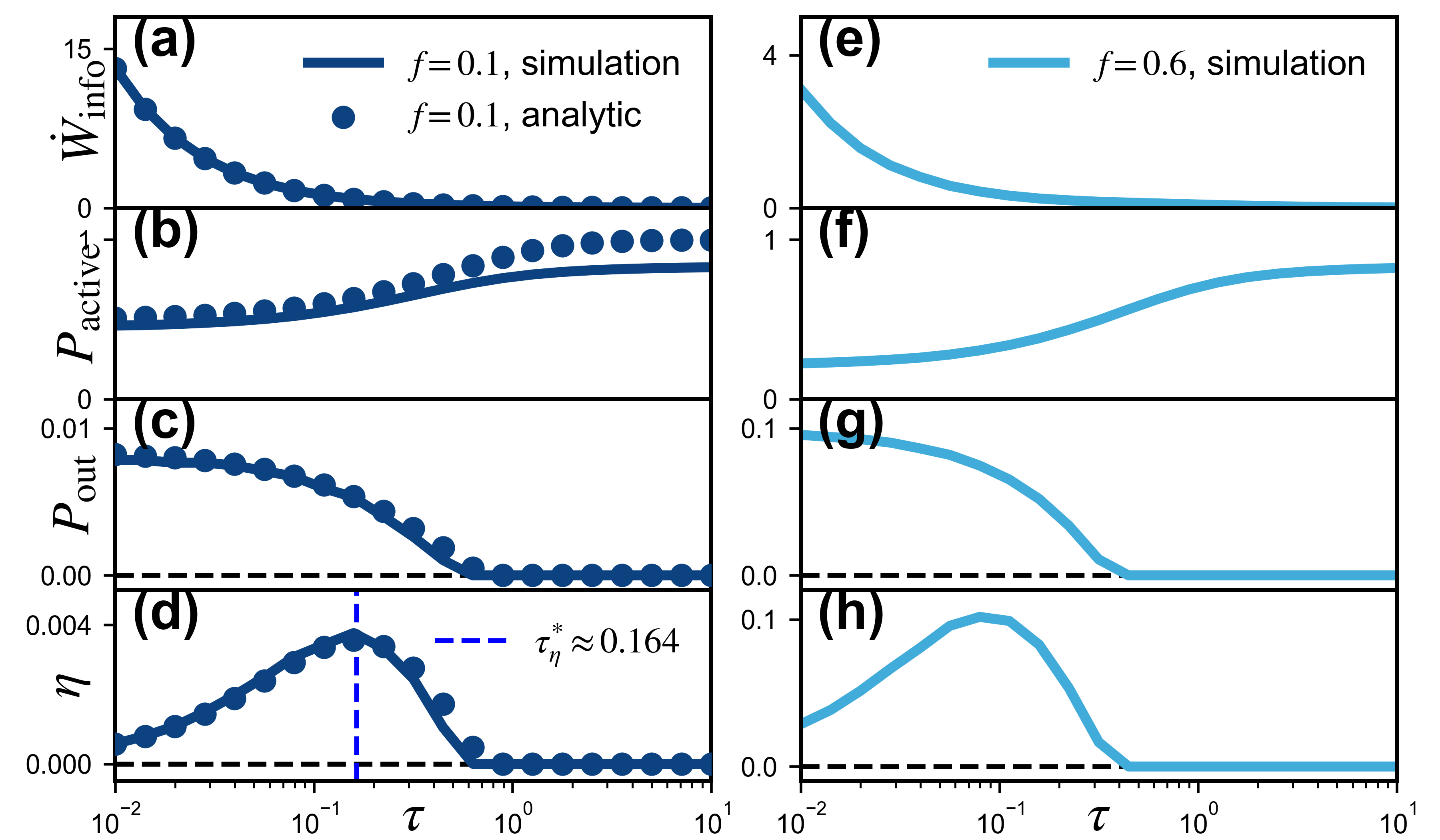}
    \caption{Energetic and information-processing performance versus measurement interval. The left column corresponds to the weak load \(f=0.1\), where solid curves show simulations and symbols denote the rapid-feedback analytical prediction. The right column shows simulations for the stronger load \(f=0.6\). (a,e) Minimum information-processing power \(\dot{W}_{\mathrm{info}}\). (b,f) Active input power \(P_{\mathrm{active}}=\langle \sigma\circ\dot{x}_1\rangle\). (c,g) Useful mechanical output power \(P_{\mathrm{out}}=\max\{0,-f\langle v\rangle\}\). (d,h) Information-inclusive efficiency \(\eta=P_{\mathrm{out}}/[P_{\mathrm{active}}+\dot{W}_{\mathrm{info}}]\). The blue dashed line indicates the optimal measurement interval predicted by the weak-load theory.
}
    \label{fig:fig3}
\end{figure}

% =====================================================
\textit{Information-inclusive efficiency}.
The reversal of the mechanical response relies on repeatedly sensing the internal state and using that information to redirect active propulsion. This introduces a fundamental tradeoff between mechanical performance and informational cost. More frequent measurements strengthen the correlation between the internal state and active propulsion, thereby enhancing motion against the applied force. This improved control, however, increases the information-processing cost per unit time due to more frequent measurements.
To quantify the energetic consequence of this tradeoff, we define the information-inclusive efficiency as $\eta=P_{\rm out}/(P_{\rm active}+\dot W_{\rm info})$, where $P_{\rm out}=\max\{0,-f\langle v\rangle\}$ is the useful mechanical output, $P_{\rm active}=\langle\sigma\circ\dot{\Tilde{x}}_1\rangle$ is the power supplied by active propulsion where $\circ$ is the Stratonovich product \cite{Sekimoto1998}, and $\dot W_{\rm info}$ is the information-processing cost rate. This definition separates two distinct resources: activity supplies the energy for mechanical work, whereas information determines how that energy is deployed. For an error-free binary measurement, the information acquired in each cycle is $\mathcal I=h(p)$. Under the fresh-memory assumption, Landauer's principle gives the minimal processing cost $W_{\rm info}=\mathcal{T}\mathcal I$ per cycle \cite{SagawaUeda2009,SagawaUeda2012} and hence the cost rate $\dot W_{\rm info}=\mathcal{T}\mathcal I/\tau$ \cite{GarciaMillanEtAl2025,SchuettlerEtAl2025}.

Figure~\ref{fig:fig3} illustrates the resulting tradeoff between mechanical performance and information-processing cost. Since the information acquired per measurement is bounded by $\mathcal I\leq\ln2$, decreasing $\tau$ drives the information-processing cost rate to diverge as $\dot W_{\rm info}\sim1/\tau$ [Figs.~\ref{fig:fig3}(a) and \ref{fig:fig3}(e)], while the output power remains finite [Figs.~\ref{fig:fig3}(c) and \ref{fig:fig3}(g)]. Consequently, the information-inclusive efficiency vanishes as $\tau\to0$ [Figs.~\ref{fig:fig3}(d) and \ref{fig:fig3}(h)]. At the opposite extreme, slow feedback incurs little information-processing cost but produces only weak response reversal and mechanical output. This tradeoff yields a finite measurement interval that maximizes the information-inclusive efficiency, which generally differs from the interval that maximizes mechanical power.
This behavior is captured analytically in the weak-load regime $f\ll 1$, where $P_{\rm active}\simeq1-\chi^{2}/2$, $P_{\rm out}\simeq(\chi\lambda-1)f^{2}/2$, and $\dot W_{\rm info}\simeq \mathcal{T}(\ln2-\lambda^{2}f^{2}/2)/\tau$ [blue dots in Figs. \ref{fig:fig3}(a)-(d), see Sec. S3]. To leading order in $f$, one obtains
\begin{equation}
    \label{eq:eta}
    \eta\approx\frac{(\chi\lambda-1)f^{2}}{2-\chi^{2}+2\mathcal{T}\ln2/\tau},
\end{equation}
for the reverse-transport regime $\chi\lambda>1$. 
In this linear-response regime, the induced velocity is proportional to the load, so the mechanical output scales as $f^2$, while the active and information-processing costs remain finite as $f\to0$. The efficiency therefore takes the form $\eta\simeq f^2\Phi(\tau)$, implying that the measurement interval maximizing $\eta$ is independent of $f$ to leading order. The optimal feedback timescale is thus set by the intrinsic balance between feedback-induced response amplification and information-processing cost, rather than by the strength of the applied load.
Interestingly, the active input power $P_{\rm active}$ decreases with increasing measurement frequency [Figs.~\ref{fig:fig3}(b) and \ref{fig:fig3}(f)]. This is because rapid feedback more strongly aligns the active orientation with the dimer extension, causing the active particle to push more persistently against the spring and thereby reducing the work performed by active propulsion. The enhanced mechanical output therefore reflects a more effective use of active energy, rather than an increase in its supply. Information supplies no energy itself, but controls how active energy is converted into motion against the applied force.

% ==================================================
\textit{Conclusion}.
We have shown that information processing can reshape nonequilibrium response by allowing an internal mechanical state to sense, store, and transform external perturbations into directed motion. An external perturbation first biases the internal configuration, triggering a self-reinforcing information–mechanical feedback that directs active propulsion against the load and ultimately produces negative mobility. Information therefore acts not as an energy source, but as a control resource that determines how internally supplied active energy is deployed. Unlike conventional routes to negative mobility based on built-in structural or dynamical rectification \cite{EichhornReimannHanggi2002,MachuraEtAl2007,Rizkallah2023}, the effective asymmetry required for response inversion is generated dynamically by the feedback between the internal state and active polarity. This mechanism is also different from feedback protocols in which the measured state merely selects a control action \cite{CocconiChen2024,SchuettlerEtAl2025,GarciaMillanEtAl2025}. Here the measured internal state is itself reshaped by the action, creating a closed loop in which information processing and mechanical response are dynamically coupled. Energetic analysis further reveals that more frequent feedback strengthens reverse transport at the expense of a growing information-processing cost, leading to distinct optima for power and information-inclusive efficiency. This reveals a general constraint of information-mediated control: optimal performance requires balancing mechanical output against processing cost by matching the feedback timescale to the system’s intrinsic dynamics.

Beyond the specific dimer considered here, these results suggest that information provides an additional constitutive ingredient that determines how nonequilibrium systems respond to external perturbations. By coupling internal sensing to active energy injection, one can in principle engineer transport properties that are adaptive rather than fixed by structure or interactions alone. Extending this framework to higher-dimensional and many-body systems may therefore enable programmable anisotropic, odd, and collective responses, with mechanical behavior determined not only by how matter is built, but also by how it acquires and acts on information.

% ========================================================
\begin{acknowledgments}
ZZ and ZY were supported by the National Natural Science Foundation of China No. 12374219, the National Key Research and Development Program of China (No. 2023YFA1407500), the 111 project (B16029), and Fujian Provincial Excellent Postdoctoral Program. SK acknowledges the support by the National Natural Science Foundation of China (Grant No. 12274098) and by the Zhejiang Key Laboratory of Soft Matter Biomedical Materials (2025ZY01036 and 2025E10072).

\end{acknowledgments}

\bibliographystyle{apsrev4-1}
\bibliography{APDimer}
\end{document}

\title{Supplemental Material for ``An Informational Route to Negative Mobility''}

\author{Ziluo Zhang}
%\thanks{Corresponding author: zhyou@xmu.edu.cn}
\affiliation{Fujian Provincial Key Laboratory for Soft Functional Materials Research,
Research Institute for Biomimetics and Soft Matter, Department of
Physics, Xiamen University, Xiamen, Fujian 361005, China}

\author{Shigeyuki Komura}
\thanks{Corresponding author: komura@wiucas.ac.cn}
\affiliation{Zhejiang Key Laboratory of Soft Matter Biomedical Materials, Wenzhou Institute, University of Chinese Academy of Sciences, Wenzhou 325001, China}

\author{Zhihong You}
\thanks{Corresponding author: zhyou@xmu.edu.cn}
\affiliation{Fujian Provincial Key Laboratory for Soft Functional Materials Research,
Research Institute for Biomimetics and Soft Matter, Department of
Physics, Xiamen University, Xiamen, Fujian 361005, China}
\date{\today}

\maketitle

\section{Numerical Methods}

In this section, we present the nondimensionalization procedure and describe the numerical methods in detail.

We nondimensionalize time by the mean run time $\alpha^{-1}$, length by the mean run length $\ell_0=u_0/\alpha$, force by $\gamma u_0$. The dimensionless particle positions $\tilde{x}_i=x_i/\ell_0$ then obey
\begin{subequations}
\label{eq:x1x2}
\begin{align}
\dot{\tilde{x}}_1 &= \sigma-\kappa \distance+\tilde{\xi}_1(t),\\
\dot{\tilde{x}}_2 &= f+\kappa \distance+\tilde{\xi}_2(t),
\end{align}
\end{subequations}
where $\distance=\tilde{x}_1-\tilde{x}_2$ is the relative separation, $\kappa=k/(\gamma\alpha)$ is the dimensionless spring stiffness, $f=F_{\mathrm{ext}}/(\gamma u_0)$ is the applied force, and $\sigma=\pm1$ is the active polarity. The dimensionless thermal noises are independent and satisfy
\begin{equation}
\label{eq:thermal-noise}
\left\langle \tilde{\xi}_i(t)\tilde{\xi}_j(t')\right\rangle
=
2\mathcal{T}\delta_{ij}\delta(t-t'),
\end{equation}
with $\mathcal{T}=\alpha k_{\mathrm B}T/(\gamma u_0^2)$ the dimensionless temperature.

Introducing the center-of-mass coordinate $X=(\tilde{x}_1+\tilde{x}_2)/2$, Eqs.~\eqref{eq:x1x2} can be converted to
\begin{subequations}
\label{eq:qX}
\begin{align}
\dot{\distance}
&=
-2\kappa\distance-f+\sigma(t)+\xi_{\distance}(t),\\
\dot{X}
&=
\frac{f+\sigma(t)}{2}+\xi_X(t),
\end{align}
\end{subequations}
where $\xi_{\distance}=\tilde{\xi}_1-\tilde{\xi}_2$ and $\xi_X=(\tilde{\xi}_1+\tilde{\xi}_2)/2$, satisfying
\begin{equation}
\left\langle\xi_{\distance}(t)\xi_{\distance}(t')\right\rangle
=
4\mathcal{T}\delta(t-t'),
\qquad
\left\langle\xi_X(t)\xi_X(t')\right\rangle
=
\mathcal{T}\delta(t-t'),
\end{equation}
with $\langle\xi_{\distance}(t)\xi_X(t')\rangle=0$. Although mathematically equivalent to Eqs. \eqref{eq:x1x2}, Eqs. \eqref{eq:qX} recast the dimer dynamics as an external translational degree of freedom $X$ coupled to an internal configurational state $q$, thereby separating the observable transport from the internal coordinate through which the force is sensed and feedback acts. For this reason, we solve Eqs. \eqref{eq:qX} numerically rather than the original particle-level equations.

The stochastic dynamics are simulated using an event-driven scheme. Immediately after each measurement at $t_n=n\tau$, we draw the waiting time $t_{\rm tum}$ to the next tumble from an exponential distribution with unit rate. The remaining time to the next measurement is updated after each tumble. If $t_{\rm tum}\geq\tau$, the polarity remains fixed throughout the measurement interval, and we propagate the coordinates over $\Delta t=\tau$ to $t_{n+1}$ before applying the feedback update. If $t_{\rm tum}<\tau$, we instead propagate the coordinates over $\Delta t=t_{\rm tum}$, reverse the polarity as $\sigma\rightarrow-\sigma$, draw a new tumble time, and repeat this procedure over the remaining portion of the measurement interval. Once the next sampled tumble would occur after $t_{n+1}$, the coordinates are advanced exactly to $t_{n+1}$, where the premeasurement configuration is recorded and $\sigma$ is reset according to the feedback rule. This construction resolves every tumble and measurement event exactly without introducing a finite integration time step.

Over any interval $\Delta t$ during which $\sigma$ remains fixed, the internal coordinate obeys an Ornstein--Uhlenbeck process with a fixed point $q_*=(\sigma-f)/(2\kappa)$ and a relaxation rate $a=2\kappa$. Its evolution can therefore be sampled exactly as 
\begin{equation}
\label{eq:ou-endpoint}
\distance(t+\Delta t) = \distance_* + [\distance(t)-\distance_*]e^{-a\Delta t} + \eta_{\distance},
\end{equation}
where $\eta_q$ is a zero-mean Gaussian variable with variance $\langle\eta_q^2\rangle=(2\mathcal{T}/a)(1-e^{-2a\Delta t})$. The external coordinate is propagated similarly from its drift-diffusion equation, giving 
\begin{equation}
X(t+\Delta t) = X(t) + \frac{f+\sigma}{2}\Delta t + \eta_X,
\end{equation}
where $\eta_X$ is Gaussian with zero mean and variance $\langle\eta_X^2\rangle=\mathcal{T}\Delta t$.

The simulations are initialized at $q(0)=0$, $X(0)=0$, and $\sigma(0)=1$. The characteristic correlation times of the internal coordinate and active polarity are $t_r=(2\kappa)^{-1}$ and $\tau_\sigma=1/2$, respectively, where the latter follows from the unit-rate Poisson flipping process. We discard the initial interval $0\leq t<50t_r$ as a transient and evaluate steady-state observables over $50t_r\leq t\leq500t_r$. The notation $\langle\cdot\rangle$ in the main text denotes an ensemble average over at least $4000$ independent realizations, each averaged over this steady-state sampling interval, with $4000$ trajectories used by default.

% ==================================================
\section{Rapid-feedback analytic theory}
In this section, we present a detailed derivation of the analytical theory in the rapid-feedback regime.

\subsection*{A. Cycle-averaged polarity memory}

The feedback protocol reset the active polarity only at the measurement events, while intrinsic tumbling progressively erases this orientational memory between updates. Immediately after the measurement at $t_n$, the polarity is set by the observed configuration, $\sigma_0=\operatorname{sgn}[\distance(t_n)]$. Thereafter, $\sigma$ evolves as a symmetric telegraph process with unit reversal rate, so its conditional mean decays as 
\begin{equation}
\label{eq:telegraph-memory}
\ave{\sigma(s)\mid\sigma(0)=\sigma_0}=\sigma_0e^{-2s}.
\end{equation}
Thus, the measured internal configuration provides the initial polarity, whereas tumbling controls how long that information remains dynamically available.

The fraction of this polarity retained over one measurement cycle is obtained by averaging the exponential decay over the interval $\tau$, giving 
\begin{equation}
\label{eq:chi-definition}
\chi(\tau)=\frac{1}{\tau}\int_0^\tau e^{-2s}  \dint s=\frac{1-e^{-2\tau}}{2\tau}.
\end{equation}
The function $\chi$ therefore quantifies the memory of the feedback protocol: $\chi\to1$ for rapid measurements, when little information is lost between updates, whereas $\chi$ decays to zero for longer intervals as tumbling randomizes the orientation. Because each cycle begins with $\sigma_0=\operatorname{sgn}[\distance(t_n)]$, averaging over measurement outcomes yields the exact relation $\langle\sigma\rangle=\chi(\tau)\zeta$, where $\zeta=\langle\operatorname{sgn}[\distance(t_n)]\rangle$ is the premeasurement configurational polarization.

This relation separates the feedback mechanism into two steps. The applied force first biases the internal configuration through $\zeta$, and the feedback converts the retained fraction $\chi\zeta$ of that bias into active propulsion. Substitution into the exact force balance gives 
\begin{equation}
\label{eq:velocity-zeta}
\ave v=\frac{1}{2}\left[f+\chi(\tau)\zeta\right].
\end{equation}
The transport response is therefore governed by the competition between the direct mechanical drive $f$ and the opposing information-induced active contribution $\chi\zeta$. For $f>0$, reverse transport occurs when the latter overscreens the applied force, namely when $\chi\zeta<-f$.

% -----------------------------------------------
\subsection*{B. Rapid-feedback stationary distribution}

For sufficiently rapid measurements, the active polarity remains closely correlated with the instantaneous sign of the separation. The feedback can then be represented by the cycle-averaged closure $\sigma\simeq\chi\sgn(\distance)$, which replaces the discrete measurement protocol by an effective configuration-dependent propulsion. The probability distribution of the internal coordinate $\rho(q)$ then obeys the Fokker–Planck equation
\begin{equation}
\label{eq:effective-fokker-planck}
\partial_t\rho=-\partial_\distance\left\{\big[-2\kappa \distance-f+\chi \sgn(\distance)\big]\rho\right\}+2\mathcal{T}\partial_\distance^2\rho.
\end{equation}
This equation makes the feedback mechanism transparent. The harmonic force confines the relative coordinate, the external load tilts the configurational landscape, and the feedback term drives the particles apart with a direction determined by the sign of $\distance$. The latter creates two preferred configurational sectors, $\distance>0$ and $\distance<0$, while the force changes their relative statistical weights.

The corresponding confined stationary distribution \cite{Risken:1989} is
\begin{equation}
\label{eq:effective-density}
\rho_{\mathrm{eff}}(\distance;f,\tau)=\frac{1}{\mathcal Z}\Exp{\left[\frac{-\kappa \distance^2-f\distance+\chi(\tau)|\distance|}{2\mathcal{T}}\right]}.
\end{equation}
The term $\chi|q|$ expresses the self-reinforcing nature of the feedback: the measured configuration determines the propulsion direction, while the resulting propulsion reshapes the configuration to increase the probability of observing the same state in future measurements. The internal coordinate therefore acts simultaneously as an information carrier and a dynamical memory. By contrast, the term $-f\distance$ favors one sign over the other and encodes the direction of the applied force. The cusp at $\distance=0$ reflects the abrupt reversal of the feedback force when the particle ordering changes. The stationary equation is therefore understood piecewise on the two half-lines, with a continuous density and vanishing probability current at the origin. No singular probability mass is generated there.

To evaluate the statistical weights of the two configurational sectors, we define
\begin{equation}
\label{eq:g-definition}
g(x)=\int_0^\infty\Exp{\frac{-\kappa \distance^2+x\distance}{2\mathcal{T}}} \dint \distance.
\end{equation}
Completing the square gives
\begin{equation}
\label{eq:g-closed-form}
g(x)=\sqrt{\frac{\pi \mathcal{T}}{2\kappa}}\Exp{\frac{x^2}{8\kappa \mathcal{T}}}\left[1+\erf\left(\frac{x}{\sqrt{8\kappa \mathcal{T}}}\right)\right],
\end{equation}
where $\erf(x)$ is the error function.
On the positive half-line, the force opposes the feedback contribution, whereas on the negative half-line the two enter with the same sign. Their respective weights are therefore
\begin{equation}
\label{eq:halfline-weights}
W_+=g(\chi-f),\qquad W_-=g(\chi+f).
\end{equation}
The probability of measuring a positive separation follows as
\begin{equation}
\label{eq:p-effective}
p(f,\tau)=\frac{g(\chi-f)}{g(\chi-f)+g(\chi+f)},
\end{equation}
and the corresponding configurational polarization is
\begin{equation}
\label{eq:zeta-effective}
\zeta(f,\tau)=\frac{g(\chi-f)-g(\chi+f)}{g(\chi-f)+g(\chi+f)}.
\end{equation}
Because $g(x)$ increases monotonically with $x$, any positive force gives $W_->W_+$ and hence $\zeta<0$. The external perturbation is thus recorded as a signed imbalance between the two configurational sectors, while feedback amplifies that imbalance by stabilizing configurations with larger $|\distance|$. These expressions generate the analytical results in Figs.~2(a)--2(c) of the main text.

\subsection*{C. No-feedback susceptibility from the stationary distribution of a trapped run-and-tumble particle}

The rapid-feedback approximation used above is designed for small measurement intervals and should not be extrapolated to the no-feedback limit. We therefore determine the reference susceptibility $\lambda_0$ independently from the exact stationary statistics of the relative coordinate in the absence of feedback. Here the argument $0$ refers to $\chi=0$, corresponding to $\tau\to\infty$, rather than to the continuous-feedback limit $\tau\to0$.

In the absence of feedback, the dimensionless relative coordinate obeys
\begin{equation}
    \dot q=\sigma-f-2\kappa q+\sqrt{4\mathcal T}\,\xi(t),
    \label{eq:no_feedback_q}
\end{equation}
where $\sigma=\pm1$ switches sign with unit rate and $\langle\xi(t)\xi(t')\rangle=\delta(t-t')$. For $f=0$, Eq.~\eqref{eq:no_feedback_q} is precisely a one-dimensional run-and-tumble particle in a harmonic potential with thermal diffusion. The corresponding stationary distribution is known exactly \cite{GarciaMillanPruessner2021}. 

A constant force $f$ only shifts the center of the harmonic confinement. Introducing
\begin{equation}
    y=q+\frac{f}{2\kappa},
\end{equation}
Eq.~\eqref{eq:no_feedback_q} becomes
\begin{equation}
    \dot y=\sigma-2\kappa y+\sqrt{4\mathcal T}\,\xi(t),
\end{equation}
which is independent of $f$. Hence the stationary distribution at finite force is simply
\begin{equation}
    P_f(q)=P_0\left(q+\frac{f}{2\kappa}\right),
    \label{eq:shifted_distribution}
\end{equation}
where $P_0(q)$ is the symmetric stationary distribution at $f=0$. The configurational polarization in the no-feedback limit is therefore
\begin{align}
    \zeta(f)
    &=
    \int_0^\infty dq\,P_f(q)
    -
    \int_{-\infty}^{0}dq\,P_f(q)
    \nonumber\\
    &=
    -2\int_0^{f/(2\kappa)}dq\,P_0(q).
    \label{eq:zeta_no_feedback}
\end{align}
Because $P_0(q)$ is even, its expansion around the origin contains only even powers of $q$, giving
\begin{equation}
    \zeta(f)
    =
    -\frac{P_0(0)}{\kappa}f
    +O(f^3).
\end{equation}
Comparing this result with the weak-load definition
\begin{equation}
    \zeta(f)=-\lambda_0f+O(f^3),
\end{equation}
we obtain the general relation
\begin{equation}
    \lambda_0=\frac{P_0(0)}{\kappa}.
    \label{eq:lambda0_P0}
\end{equation}

We now specialize to the value $\kappa=1/2$ used throughout this work.  The exact stationary distribution at finite $D=2\mathcal{T}$ is given by the Hermite-series solution of Ref.~\cite{GarciaMillanPruessner2021}. Evaluating this distribution at the origin, only the even terms $n=2m$ contribute, giving
\begin{equation}
P_0(0)=\frac{1}{\sqrt{2\pi D}}\sum_{m=0}^{\infty}\frac{(-1)^m}{m!(2m+1)(2D)^m}.
\label{eq:P0_series}
\end{equation}

The series in Eq.~\eqref{eq:P0_series} can be summed using the expansion
\begin{equation}
\operatorname{erf}(x)=\frac{2}{\sqrt{\pi}}\sum_{m=0}^{\infty}\frac{(-1)^m x^{2m+1}}{m!(2m+1)}.
\end{equation}
Setting $x=1/\sqrt{2D}$ gives
\begin{equation}
P_0(0)=\frac{1}{2}\operatorname{erf}\left(\frac{1}{\sqrt{2D}}\right).
\end{equation}
Since $D=2\mathcal T$, the no-feedback susceptibility therefore becomes
\begin{equation}
\lambda_0=\operatorname{erf}\left(\frac{1}{2\sqrt{\mathcal T}}\right), \ \text{for} \ \kappa=1/2.
\label{eq:lambda0_exact}
\end{equation}
For $\mathcal T=0.2$, this gives $\lambda_0\simeq0.886$. This result follows from the exact stationary distribution of the no-feedback run-and-tumble dynamics and therefore does not rely on the rapid-feedback approximation. This closed-form expression applies specifically to $\kappa=1/2$. For general $\kappa$, $\lambda_0$ can be obtained either numerically from the exact stationary distribution in \cite{GarciaMillanPruessner2021}  or analytically in terms of a confluent hypergeometric function.

% -----------------------------------------------
\subsection*{D. Weak-load susceptibility and feedback gain}

The stationary distribution derived above describes the response at finite load, but the onset of absolute negative mobility is controlled by the linear response to an arbitrarily weak perturbation. We therefore examine how a small force first biases the measured configuration and how feedback subsequently converts this bias into an opposing active polarity. The central quantity is the sensitivity of the positive-separation probability $p$ to the applied force.

To characterize this sensitivity, we introduce the logarithmic derivative
\begin{equation}
\label{eq:ell-definition}
j(x)=\frac{g'(x)}{g(x)}.
\end{equation}
Differentiating Eq.~\eqref{eq:p-effective} with respect to the load gives
\begin{equation}
\label{eq:p-load-derivative}
\partial_f p=-p(1-p)\left[j(\chi-f)+j(\chi+f)\right].
\end{equation}
At zero load, inversion symmetry makes the two signs of the separation equally probable, so that $p(0,\tau)=1/2$ and $\zeta(0,\tau)=0$. We define the weak-load configurational susceptibility as
\begin{equation}
\label{eq:lambda-definition}
\lambda(\tau)=j[\chi(\tau)]=\frac{g'[\chi(\tau)]}{g[\chi(\tau)]}.
\end{equation}
The quantity $\lambda$ measures how strongly an infinitesimal force changes the relative statistical weights of the two configurational sectors. Because the effective distribution itself depends on the feedback-retention factor $\chi$, $\lambda$ is a closed-loop susceptibility: feedback not only reads the configurational bias but also modifies how sensitively that bias responds to the load.

Expanding around $f=0$ gives
\begin{subequations}
\label{eq:weak-load-expansions}
\begin{align}
p(f,\tau)&=\frac{1}{2}-\frac{1}{2}\lambda(\tau)f+O(f^3),\\
\zeta(f,\tau)&=-\lambda(\tau)f+O(f^3),\\
\ave{\sigma}&=-\chi(\tau)\lambda(\tau)f+O(f^3).
\end{align}
\end{subequations}
Only odd powers of $f$ appear in the polarization because inversion symmetry requires $p(-f,\tau)=1-p(f,\tau)$ and $\zeta(-f,\tau)=-\zeta(f,\tau)$. For a positive force, $\zeta<0$: the passive particle is displaced preferentially to the right of the active particle, and feedback converts this configurational preference into a negative mean active polarity.

Substituting Eq.~\eqref{eq:weak-load-expansions} into the exact force balance gives
\begin{equation}
\label{eq:weak-load-velocity}
\ave{v}=\frac{f}{2}\left[1-\chi(\tau)\lambda(\tau)\right]+O(f^3),
\end{equation}
and hence the weak-load mobility
\begin{equation}
\label{eq:zero-load-mobility}
\mu_0(\tau)=\left.\frac{\partial\ave{v}}{\partial f}\right|_{f=0}=\frac{1}{2}\left[1-\chi(\tau)\lambda(\tau)\right].
\end{equation}
This result separates the response into a bare mechanical contribution, represented by the first term, and an opposing information-mediated active contribution, represented by $\chi\lambda$. It is therefore natural to define the feedback gain
\begin{equation}
\label{eq:feedback-gain}
G_{\mathrm{fb}}(\tau)=\chi(\tau)\lambda(\tau).
\end{equation}
For $G_{\mathrm{fb}}<1$, feedback partially screens the ordinary positive response but does not reverse it. At $G_{\mathrm{fb}}=1$, the active and mechanical contributions exactly balance and the zero-load mobility vanishes. For
\begin{equation}
G_{\mathrm{fb}}>1,
\end{equation}
the feedback-induced active response exceeds the bare drift, producing active overscreening and absolute negative mobility.

% -----------------------------------------------
\subsection*{E. Information-encoding interpretation}

The configurational susceptibility can also be expressed in information-theoretic terms. Each measurement reduces the continuous internal coordinate $\distance$ to a binary record specifying whether $\distance$ is positive or negative. At zero load, the two outcomes are equally likely and the record is maximally uncertain. An applied force breaks this balance, making one outcome more probable and thereby leaving a statistically detectable signature in the internal configuration.

The Shannon entropy of the measured sign is
\begin{equation}
\label{eq:binary-entropy}
h(p)=-p\ln p-(1-p)\ln(1-p).
\end{equation}
Its maximum value is $\ln 2$, attained at $p=1/2$. We therefore define the load-induced entropy deficit as
\begin{equation}
\label{eq:entropy-deficit}
\Delta h=\ln 2-h(p).
\end{equation}
This quantity measures how much the applied load reduces the uncertainty of the configurational bit. It should not be confused with the energetic cost of processing information considered below: here, $\Delta h$ is used only as a statistical measure of how clearly the internal state encodes the perturbation.

Using $p=(1+\zeta)/2$, the entropy deficit can be written directly in terms of the configurational polarization,
\begin{equation}
\label{eq:entropy-deficit-zeta}
\Delta h=\frac{1}{2}\left[(1+\zeta)\ln(1+\zeta)+(1-\zeta)\ln(1-\zeta)\right].
\end{equation}
For a weak configurational bias,
\begin{equation}
\label{eq:entropy-deficit-expansion}
\Delta h=\frac{\zeta^2}{2}+\frac{\zeta^4}{12}+O(\zeta^6).
\end{equation}
The leading contribution is quadratic because the entropy does not distinguish the sign of the bias: forces of equal magnitude and opposite direction reduce the uncertainty by the same amount.

Combining this expansion with $\zeta=-\lambda f+O(f^3)$ gives
\begin{equation}
\label{eq:lambda-information}
\Delta h=\frac{\lambda^2f^2}{2}+O(f^4),
\qquad
\lambda=\lim_{f\to0}\frac{\sqrt{2\Delta h}}{|f|}.
\end{equation}
Thus, $\lambda$ quantifies how rapidly a weak mechanical perturbation becomes encoded as a statistically distinguishable binary internal state. The factor $\chi$ then quantifies how much of this measured record survives intrinsic tumbling and remains available to generate active propulsion. The feedback gain $G_{\mathrm{fb}}=\chi\lambda$ therefore combines information encoding and dynamical memory into a single criterion for response reversal.

% =====================================================

\section{Energetics and information-inclusive efficiency}

In this section, we examine how the feedback-controlled dimer converts active energy into useful mechanical work within the framework of stochastic energetics \cite{Sekimoto1998,Seifert2012,Speck2016}. We first identify the mechanical output and the power supplied by self-propulsion, and then assign a minimum thermodynamic cost to the repeated processing of the measurement record \cite{SagawaUeda2012}. Combining these contributions yields an information-inclusive efficiency that exposes the competition between stronger control and the energetic cost of rapid feedback, complementing recent formulations of active and information-powered engines \cite{CocconiChen2024,GarciaMillanEtAl2025,SchuettlerEtAl2025}.

\subsection*{A. Mechanical output}

Mechanical work is extracted only when the dimer moves against the applied force. For $f>0$, ordinary motion with $\ave{v}>0$ corresponds to the external force doing positive work on the system and therefore does not constitute useful output. By contrast, reverse transport with $\ave{v}<0$ allows the system to perform work against the load, as in active engines that convert self-propulsion into mechanical work \cite{Speck2022,GarciaMillanEtAl2025,SchuettlerEtAl2025}. We consequently define the useful output power as
\begin{equation}
\label{eq:output-power-definition}
P_{\mathrm{out}}=\pos{-f\ave{v}},
\end{equation}
where $\pos{x}=\max(0,x)$ ensures that the output is non-negative and vanishes outside the engine regime.

Using the rapid-feedback relation in Eq.~\eqref{eq:velocity-zeta}, the output becomes
\begin{equation}
\label{eq:output-power-effective}
P_{\mathrm{out}}^{\mathrm{eff}}=\pos{\frac{f}{2}\left[-f-\chi\zeta(f,\tau)\right]}.
\end{equation}
The two terms inside the brackets have distinct physical origins. The term $-f$ represents the direct mechanical tendency to move along the applied force, whereas $-\chi\zeta$ represents the opposing active polarity generated by feedback. Positive output is possible only when the information-mediated contribution is sufficiently large to overcome the bare mechanical drift.

In the weak-load reverse-transport regime, substituting Eq.~\eqref{eq:weak-load-expansions} gives
\begin{equation}
\label{eq:output-power-weak}
P_{\mathrm{out}}=\frac{1}{2}\left(\chi\lambda-1\right)f^2+O(f^4).
\end{equation}
The condition $\chi\lambda>1$ is therefore simultaneously the condition for negative mobility and for positive mechanical output at weak load. The output is quadratic in $f$ because both the opposing velocity and the load against which work is performed are linear in the perturbation. As $\tau\to0$, the factors $\chi$ and $\lambda$ approach finite limits, so the output power saturates rather than diverges in the continuous-feedback limit.

% -----------------------------------------------
\subsection*{B. Active input power}

The energy required to move against the load is ultimately supplied by self-propulsion. Feedback does not act as an independent energy source. Instead, it controls the direction in which the active energy is deployed. Following stochastic energetics, mechanical work is defined along the stochastic trajectory through the force-displacement product \cite{Sekimoto1998,Seifert2012}. Treating self-propulsion as an effective nonconservative force provides the corresponding active-work functional \cite{Speck2016}. The dimensionless mechanical power injected by the active force is therefore
\begin{equation}
\label{eq:active-power-definition}
P_{\mathrm{active}}=\ave{\sigma\circ\dot{\tilde{x}}_1},
\end{equation}
where $\circ$ denotes the Stratonovich product. The polarity is piecewise constant, and the particle positions remain continuous at both tumble and feedback events. Consequently, the polarity has no quadratic covariation with the particle position, and the It\^o and Stratonovich conventions give the same result for this active-work functional.

Using the active-particle equation of motion, $\dot{\tilde{x}}_1=\sigma-\kappa\distance+\xi_1$, together with $\sigma^2=1$, we obtain
\begin{equation}
\label{eq:active-power-identity}
P_{\mathrm{active}}=1-\kappa\ave{\sigma\distance}.
\end{equation}
The correlation $\ave{\sigma\distance}$ measures how the propulsion direction is aligned with the spring extension. Under the feedback rule, the active particle usually points away from its passive partner, so $\sigma\distance>0$. The spring then opposes the active motion, reducing the velocity component along the propulsion direction and hence lowering the mechanical power injected by activity. Stronger feedback can therefore increase the useful output even while decreasing the active input power, because it uses the available active energy more coherently.

In the simulations, the active-extension correlation is evaluated as the continuous-time average
\begin{equation}
\label{eq:sigma-distance-numerical}
\ave{\sigma\distance}=\frac{1}{N_{\mathrm{traj}}t_{\mathrm{samp}}}\sum_{j=1}^{N_{\mathrm{traj}}}\int_0^{t_{\mathrm{samp}}}\sigma_j(t)\distance_j(t)\,\dint t,
\end{equation}
where $\distance=\tilde{x}_1-\tilde{x}_2$ and $t_{\mathrm{samp}}$ is the steady-state sampling time. Each trajectory is divided into event intervals bounded by consecutive tumble or measurement events. Within each interval, the polarity remains constant and the relative coordinate follows an Ornstein--Uhlenbeck process.

For an event interval $[t,t+\Delta t]$, we define the integrated separation
\begin{equation}
\label{eq:integrated-separation}
\mathcal{J}_{\distance}=\int_t^{t+\Delta t}\distance(s)\,\dint s.
\end{equation}
The endpoint $\distance(t+\Delta t)$ and $\mathcal{J}_{\distance}$ are sampled jointly from their exact Gaussian distribution. Since $\sigma$ is constant within the interval, its contribution to the active-extension integral is simply $\sigma\mathcal{J}_{\distance}$. Summing these contributions over all event intervals eliminates the need for a finite numerical time step and preserves the correlation between the endpoint and the integrated separation generated by the same thermal-noise realization.

Within the rapid-feedback closure, the polarity is approximated by $\sigma\simeq\chi\operatorname{sgn}\distance$, giving
\begin{equation}
\label{eq:sigma-z-closure}
\ave{\sigma\distance}\simeq\chi\ave{|\distance|}_{\mathrm{eff}}.
\end{equation}
The absolute first moment of the effective distribution in Eq.~\eqref{eq:effective-density} is
\begin{equation}
\label{eq:absolute-extension}
\ave{|\distance|}_{\mathrm{eff}}=2\mathcal{T}\left\{p\,j(\chi-f)+(1-p)\,j(\chi+f)\right\}.
\end{equation}
The corresponding finite-load estimate of the active input is therefore
\begin{equation}
\label{eq:active-power-effective}
P_{\mathrm{active}}^{\mathrm{eff}}=1-\kappa\chi\ave{|\distance|}_{\mathrm{eff}}.
\end{equation}

At weak load and sufficiently low temperature, the effective distribution is concentrated near the two preferred extensions $\distance\simeq\pm\chi/(2\kappa)$. The typical absolute extension is then
\begin{equation}
\label{eq:absolute-extension-lowT}
\ave{|\distance|}_{\mathrm{eff}}\simeq\frac{\chi}{2\kappa},
\end{equation}
which gives
\begin{equation}
\label{eq:active-power-weak}
P_{\mathrm{active}}\simeq1-\frac{\chi^2}{2}.
\end{equation}
This expression shows directly why rapid feedback lowers the active input: increasing $\chi$ strengthens the alignment between propulsion and spring extension, so a larger fraction of the active force is spent working against the internal restoring force.

% -----------------------------------------------
\subsection*{C. Information-processing cost}

The feedback protocol also requires a controller to measure the sign of the internal separation, store the result, and use it to reset the active polarity. The thermodynamic role of measurements, memories, and feedback is described within information thermodynamics, where the acquired information modifies the energetic bounds on feedback-controlled processes \cite{SagawaUeda2009,SagawaUeda2012,ParrondoHorowitzSagawa2015}. For an error-free measurement, the controller memory contains a binary record $M=\operatorname{sgn}\distance$. If the probability of the positive outcome is $p$, the Shannon entropy of this record is
\begin{equation}
I=H(M)=h(p)=-p\ln p-(1-p)\ln(1-p),
\end{equation}
measured in natural units.

We consider a fresh-memory implementation in which one memory element is used during each feedback cycle and reset before being reused. Under the Landauer bound, erasing a memory record with entropy $h(p)$ requires at least $\mathcal{T}h(p)$ of dimensionless work per cycle \cite{Landauer1961,Bennett1982,SagawaUeda2009}. Since measurements are performed once every interval $\tau$, the corresponding minimum information-processing power is
\begin{equation}
\label{eq:information-power}
\dot{W}_{\mathrm{info}}=\frac{\mathcal{T}h(p)}{\tau}.
\end{equation}
This expression represents an ideal lower bound for the specified fresh-memory architecture rather than the complete microscopic energetic cost of a particular controller. Alternative autonomous descriptions introduce an explicit measurement degree of freedom and quantify the controller cost through its additional entropy production \cite{CocconiChen2024}.

The dependence on $p$ and $\tau$ reflects two competing effects. A stronger load makes one measurement outcome more predictable and therefore decreases the entropy stored in each record. More rapid feedback, however, increases the number of records that must be processed per unit time. In the weak-load regime, Eqs.~\eqref{eq:weak-load-expansions} and \eqref{eq:entropy-deficit-expansion} give
\begin{equation}
\label{eq:binary-entropy-weak}
h(p)=\ln 2-\frac{\lambda^2f^2}{2}+O(f^4),
\end{equation}
and hence
\begin{equation}
\label{eq:information-power-weak}
\dot{W}_{\mathrm{info}}\simeq\frac{\mathcal{T}}{\tau}\left(\ln 2-\frac{\lambda^2f^2}{2}\right).
\end{equation}
The information contained in one measurement remains finite as $\tau\to0$, but the number of measurements per unit time grows as $1/\tau$. Consequently, the fresh-memory processing power diverges in the continuous-feedback limit even though the mechanical output remains finite.

% -----------------------------------------------
\subsection*{D. Weak-load efficiency}

The engine uses two distinct resources. Activity supplies the mechanical energy, while information processing determines how effectively that energy is directed against the external load. To account for both contributions, we define the information-inclusive efficiency as
\begin{equation}
\label{eq:efficiency-definition}
\eta=\frac{P_{\mathrm{out}}}{P_{\mathrm{active}}+\dot{W}_{\mathrm{info}}}.
\end{equation}
This quantity compares the useful mechanical output with the sum of the active mechanical input and the Landauer-limited information-processing power. It is therefore an efficiency associated with the idealized resource accounting adopted here, rather than a complete microscopic efficiency of an explicit controller.

In the weak-load reverse-transport regime, the output begins at order $f^2$, whereas the leading active and information-processing inputs remain finite at $f=0$. Retaining the leading output and the load-independent parts of the denominator gives
\begin{equation}
\label{eq:efficiency-weak}
\eta\simeq\frac{(\chi\lambda-1)f^2}{2-\chi^2+2\mathcal{T}\ln 2/\tau},
\qquad
\chi\lambda>1.
\end{equation}

Since $f^2$ appears only as an overall prefactor in Eq.~\eqref{eq:efficiency-weak}, the optimal measurement interval in the weak-load regime is determined by maximizing
\begin{equation}
\mathcal{E}(\tau)=
\frac{\chi(\tau)\lambda(\tau)-1}
{2-\chi^2(\tau)+2\mathcal{T}\ln 2/\tau},
\end{equation}
within the reverse-transport region $\chi\lambda>1$. The optimal interval $\tau_\eta^\ast$ therefore satisfies
\begin{equation}
\left.
\frac{\mathrm d}{\mathrm d\tau}\mathcal{E}(\tau)
\right|_{\tau=\tau_\eta^\ast}=0,
\end{equation}
or, equivalently,
\begin{equation}
\left.
\frac{(\chi\lambda)'}{\chi\lambda-1}
\right|_{\tau_\eta^\ast}
=
\left.
\frac{-2\chi\chi'-2\mathcal{T}\ln 2/\tau^2}
{2-\chi^2+2\mathcal{T}\ln 2/\tau}
\right|_{\tau_\eta^\ast}.
\end{equation}
For the parameters used in the main text, $\kappa=0.5$ and $\mathcal{T}=0.2$, numerical solution gives
\begin{equation}
\tau_\eta^\ast\simeq 0.164,
\qquad
\chi(\tau_\eta^\ast)\simeq0.853,
\qquad
\lambda(\tau_\eta^\ast)\simeq2.411,
\end{equation}
so that $\chi\lambda\simeq2.057$. The corresponding maximum efficiency is
\begin{equation}
\eta_{\max}\simeq0.356\,f^2.
\end{equation}
For $f=0.1$, this gives $\eta_{\max}\simeq3.56\times10^{-3}$. Thus, to leading order in the load, the optimal measurement interval is independent of $f$, whereas the maximum efficiency grows quadratically with the applied force.

The numerator measures the excess information-mediated response beyond the bare mechanical drift. The denominator contains two qualitatively different costs: the active input decreases as feedback strengthens the active--spring correlation, whereas the information-processing cost increases as measurements become more frequent.

This competition explains why the measurement interval that maximizes efficiency need not coincide with the interval that maximizes output power. For slow feedback, orientational memory is lost between measurements, the gain $\chi\lambda$ becomes too small, and little or no work is extracted. For extremely rapid feedback, the mechanical output saturates, but the fresh-memory processing cost grows without bound. The efficiency must therefore vanish in the continuous-feedback limit and may attain its maximum at a finite measurement interval.

Indeed, for $\tau\to0$,
\begin{equation}
\label{eq:chi-small-tau}
\chi(\tau)=1-\tau+\frac{2}{3}\tau^2+O(\tau^3),
\end{equation}
while $P_{\mathrm{out}}$ and $P_{\mathrm{active}}$ approach finite values. Since $\dot{W}_{\mathrm{info}}\sim\mathcal{T}h[p(0)]/\tau$, the leading small-$\tau$ behavior is
\begin{equation}
\label{eq:efficiency-small-tau}
\eta\sim\frac{P_{\mathrm{out}}(0)}{\mathcal{T}h[p(0)]}\tau,
\qquad
\tau\to0.
\end{equation}
Thus, arbitrarily frequent measurements can maximize control over the mechanical response, but they do not maximize the information-inclusive efficiency.

% -----------------------------------------------

\bibliographystyle{apsrev4-1}
\bibliography{APDimer}